      [1999/12/01 v1.4c Il Nuovo Cimento]
\documentclass{cimento}

\usepackage{graphicx}  
\title{The physics of GRB jets and their interaction with the 
progenitor star}
\author{Davide Lazzati\from{ins:x},
Brian J. Morsony\from{ins:x}, \atque
Mitchell C. Begelman\from{ins:x}}
\instlist{\inst{ins:x} JILA, Iniversity of Colorado, 440 UCB, Boulder, 
CO 80309-0440}
\PACSes{\PACSit{98.76.Rz - 98.38.j - 95.30.Dr}{}}
\begin{document}

\maketitle

\begin{abstract}
It is now generally accepted that long gamma-ray bursts are associated
with the final evolutionary stages of massive stars. As a consequence,
their jets must propagate through the stellar progenitor and break out
on their surface, before they can reach the photospheric radius and
produce the gamma-ray photons. We investigate the role of the
progenitor star in shaping the jet properties. We show that even a jet
powered by a steady engine can develop a rich phenomenology at the
stellar surface. We present special-relativistic simulations and
compare the results to analytic considerations. We show that the jet
is complex in the time as well as in the angular domain, so that
observers located along different lines of sight detect significantly
different bursts.
\end{abstract}

\section{Introduction}

Long duration gamma-ray bursts (GRBs) are likely produced by the
dissipation of bulk kinetic energy in a relativistic flow with a
Lorentz factor of at least $100$. The presence of supernova features
in the afterglow of several events\cite{stanek03} implies that the
outflow has to propagate through a stripped massive star at the end of
its evolution before producing the gamma-rays we observe. There are
therefore three mechanisms playing a role in shaping the GRB
light curves: the inner engine, the hydrodynamic interaction with the
progenitor star and the dissipation/radiative mechanism. Which of
these is the dominant one is poorly understood.

We present high resolution numerical simulations of relativistic jets
propagating through the core of massive stars\cite{morsony06}. The
simulations, performed with the adaptive mesh refinement (AMR) special
relativistic hydro code FLASH, are tailored to shed light on the
second question above: what properties in the outflow are imprinted by
its propagation through the star. To this aim, we inject a flow with
constant properties in the center of the star, we propagate it, and we
study the phenomenology on the jet emerging on the stellar surface.

\begin{figure}
\includegraphics[width=\textwidth]{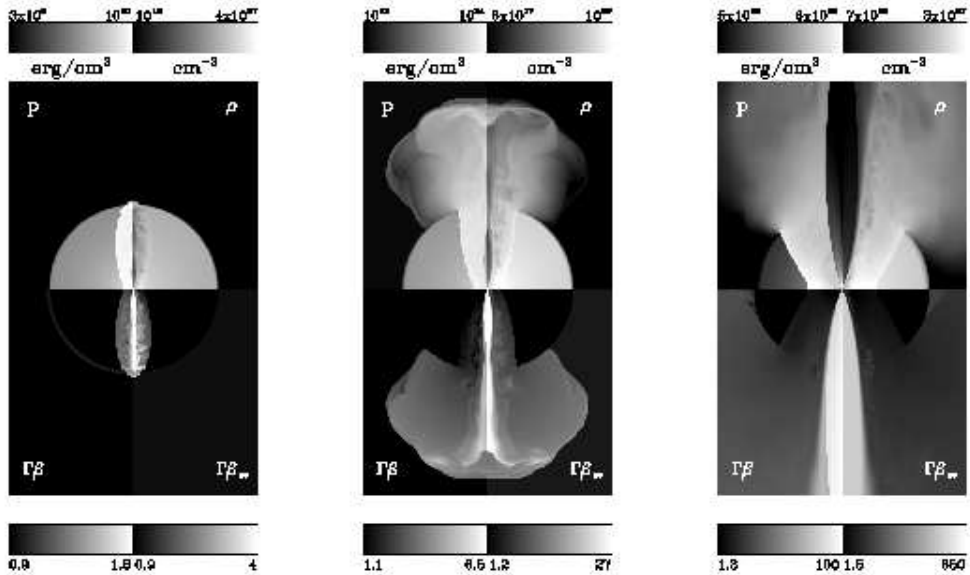} 
\caption{{Stills from a simulation of a relativistic jet with
$\theta_0=10^\circ$, $L=2.66\times10^{50}$~erg~s$^{-1}$, and
$\Gamma_0=5$. The jet propagates through a star with $R=10^{11}$~cm, a
power-law density profile and a mass of $M=15M_\odot$. Each of the
three panels is divided into four sub-panels, each showing a different
quantity. Starting from the upper right panel, in clockwise order,
panels show the density, the Lorentz factor achievable at infinity,
the actual Lorentz factor and the pressure. The gray-scale is always
logarithmic. The first panel shows a still at 10.3 seconds after the
moment at which the engine is turned on. The middle panel shows a
still at 15 seconds, while the right panel shows a still at 40
seconds. The jet is in the shocked phase in the central panel and in
the un-shocked phase in the right panel.}
\label{fig:phases}}
\end{figure}

\section{Phases}
The development of the relativistic flow is shown in the simulations
to proceed through four subsequent phases, three of which are
radiative and can, potentially, be observed in the GRB light curves.
Initially (see also the left panel of Fig.~\ref{fig:phases}), the jet
is bounded inside the star. Its head propagates subrelativistically,
yet highly supersonically, creating a bow shock that recycles shocked
hot jet material in a cocoon that surrounds the jet itself. The cocoon
pressure recollimates the jet, that propagates through the star with
an opening angle much smaller than the one with which it was injected
(10 degrees in our simulations).

With a propagation speed of several tens of per cent the speed of
light, the head of the jet reaches the surface of the star $5$ to $10$
seconds after the jet is initiated (left panel of
Fig.~\ref{fig:phases}). At this point, the cocoon energy is
released\cite{ramirez02}. The cocoon is made of hot jet material,
partially mixed through vortex shedding with the stellar material. It
does not have relativistic bulk motion, but has a potential adiabatic
acceleration up to Lorentz factors of several tens. Since it is
released through a nozzle on the stellar surface it is almost
isotropic.

\begin{figure}[!t]
\centerline{\includegraphics[width=0.75\textwidth]{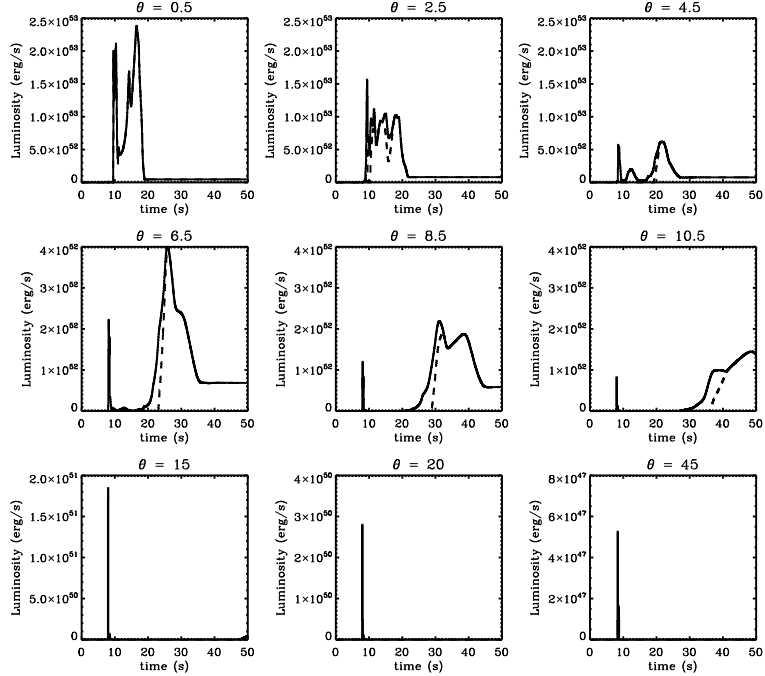}}
\caption{{Light curves of the kinetic luminosity emerging out of 
the stellar surface for observers located at different angles with
respect to the jet axis. Solid lines show the kinetic luminosity of
all the material with $\Gamma_\infty\ge2$, where $\Gamma_\infty$ is
the maximum achievable Lorentz factor in an adiabatic expansion. For
comparison, dashed lines show the luminosity of the material with
$\Gamma_\infty\ge100$. The data are from simulation
16TIg5\cite{morsony06} that has $\theta_0=10^\circ$ and
$\Gamma_0=5$. The flattening of the light curves at late times
(especially evident in the $\theta=6.5^\circ$ and $\theta=8.5^\circ$
cases) is due to the unshocked phase, when the jet enters a steady
state.}
\label{fig:lcurg5}}
\end{figure}

Simultaneously to the release of the cocoon, the relativistic jet
emerges out of the star and propagates. It has already achieved large,
albeit unsteady, values of Lorentz factor, and so it stays
collimated. In this phase, that we call the shocked jet phase, the jet
properties are severely affected by the struggle to propagate through
a cold and dense stellar core (central panel of
Fig.~\ref{fig:phases}). The jet material has been repeatedly shocked
and is characterized by alternated high-density/low-$\Gamma$ and
low-density/high-$\Gamma$ regions and by the presence of lateral
shocks. The opening angle fluctuates in this phases, but does not have
a regular increase or decrease, always being limited to several
degrees. This phase lasts few tens of seconds, depending on the
stellar properties.

Eventually, as the cocoon pressure wanes, the jet settles into a more
stable condition, in which the core of the jet expands freely until
impacts a shear layer that separates it from the stellar material
(right panel of Fig.~\ref{fig:phases}). The free-jet/shear layer are
separated by a strong shock. In this phase the jet opening angle
widens\cite{lazzati05}, until the limiting opening angle
$\theta_{\rm{lim}}=\theta_0+\Gamma_0^{-1}$ is reached, where the right
hand quantities are the injection values of opening angle and Lorentz
factor. Very small temporal variability is given to the flow by the
propagation through the star in this phase. It is therefore possible
that variability at late stages, if detected, would give us insight in
the properties of the inner engine.

\section{Light curves}

\begin{figure}[!t]
\centerline{\includegraphics[width=0.75\textwidth]{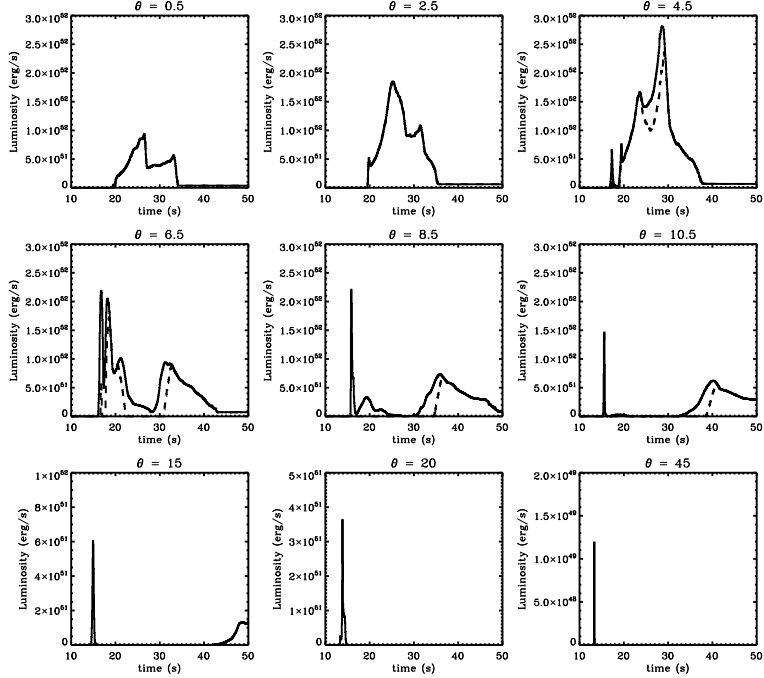}}
\caption{{Same as Fig.~\ref{fig:lcurg5} but for data from the 
simulation 16TIg2, that has a lower initial Lorentz factor
$\Gamma_0=2$.}
\label{fig:lcurg2}}
\end{figure}

Due to the complex evolution of the jet propagation inside and outside
of the star, different observers will detect different bursts. In this
section we discuss the light curves obtained for different observers
as a function of the off-axis angle. Since neither the dissipation
mechanism nor the radiative process that produce the final GRB light
curves are known, we discuss here the kinetic luminosity of the
outflow. In addition we compute our light curves just outside the
stellar surface, even though the interaction between different parts
of the flow is not over yet and can lead to some modifications at the
radiative radius. Light curves for different viewing angles and
initial conditions are shown in Fig.~\ref{fig:lcurg5}
and~\ref{fig:lcurg2}. In each figure we show the light curve of highly
relativistic material ($\Gamma\ge100$), capable of producing
$\gamma$-rays, and of material that is at least mildly relativistic
($\Gamma\ge2$), more relevant to X-ray and afterglow components.

\subsection{The large-angle observer}

Consider an observer located far from the jet axis, at an angle
$\theta>\theta_0+\Gamma_0^{-1}$. Due to the above considerations the
jet will never expand wide enough so that the observer could see it
directly. The cocoon release results however in a much wider outflow
and cocoon emission would be detected along large-angle lines of
sight. Due to its wide angle and limited energetics, cocoon emission
is rather weak, as can be seen in the bottom line of
Fig.~\ref{fig:lcurg5} and~\ref{fig:lcurg2}. Events observed under such
conditions could be GRB~980425 and/or GRB031203. The lack of detection
in the radio band of off-axis jet emission is however difficult to
account for, at least in the first case\cite{waxman04}.

\subsection{The intermediate observer}

Consider now an intermediate observer. Its viewing angle is located
outside the jet opening angle during the shocked jet phase, but inside
the limiting opening angle of the un-shocked phase. This observer will
initially detect the cocoon emission, that will be identified as a
weak precursor\cite{lazzati05a}. During the shocked jet phase, the
outflow is beamed inside a small opening angle and our observer does
not detect any emission. Eventually, as the jet opening angle spreads
to include the line of sight, the main GRB emission is detected.
Initially, the emission is due to the bright shear layer. Eventually,
the un-shocked jet emission is seen, characterized by a steady, weak
luminosity. It is during this phase, probably elusive and hard to
detect, that the inner engine is seen ``directly''.

\subsection{The on-axis observer}

Consider finally an observer located within the opening angle of the
shocked jet phase. This observer will detect radiation at all
stages. Initially she/he will detect the cocoon energy, followed,
without any interruption or dead time, by the shocked jet
emission. This is the brightest phase, since all the jet energy is
concentrated in the small opening angle of the shocked jet (see the
top panels of Fig.~\ref{fig:lcurg5} and ~\ref{fig:lcurg2}). Strong
variability, especially in the highly relativistic component
($\Gamma\ge100$, dashed line in the figures), is present in this
phase, due to the jet-cocoon interaction. After the shocked jet phase,
the light curve sharply transitions to the unshocked jet.

\bigskip

\begin{figure}
\parbox{0.49\textwidth}{
\includegraphics[width=0.48\textwidth]{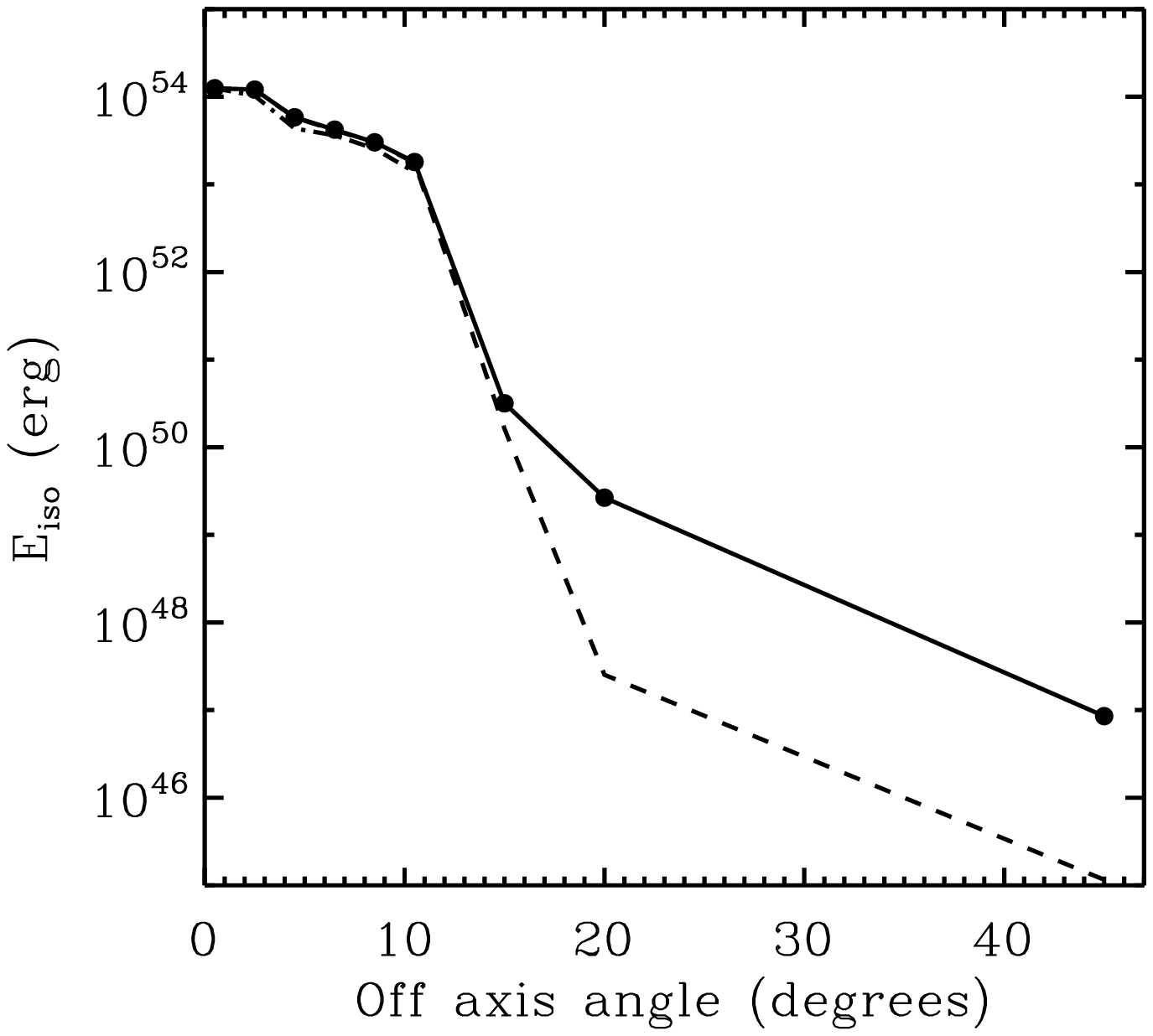}}
\parbox{0.49\textwidth}{
\includegraphics[width=0.48\textwidth]{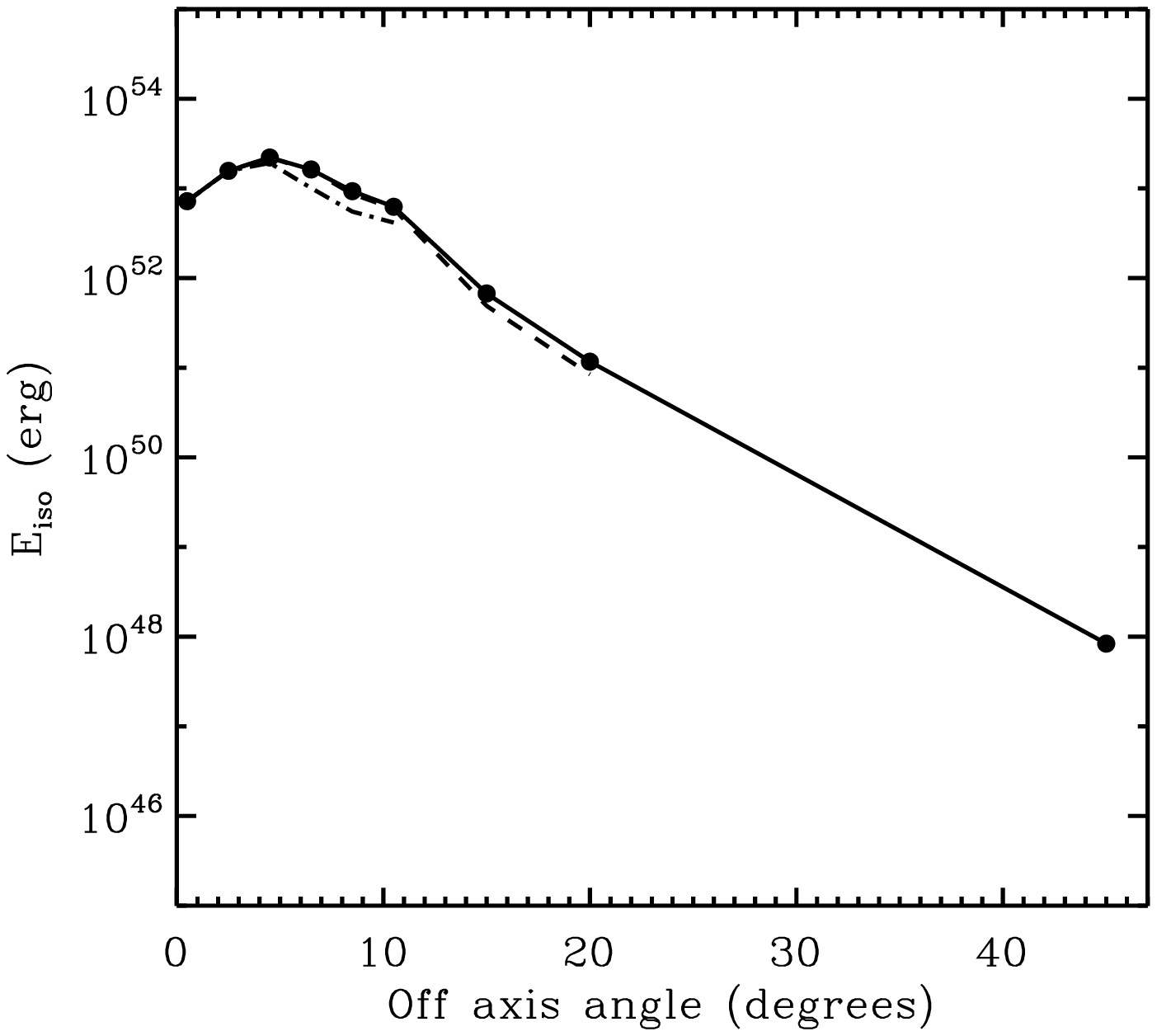}}
\caption{{Isotropic equivalent energy versus off-axis angle for the 
light curves in Fig.~\ref{fig:lcurg5} and~\ref{fig:lcurg2}. The solid
line shows all the energy of material with $\Gamma_0\ge2$ while the
dashed line shows only the energy of material with $\Gamma_0\ge100$.}
\label{fig:etet}}
\end{figure}

The comparison of the various panels of Fig.~\ref{fig:lcurg5}
and~\ref{fig:lcurg2} shows that the brightest light curves are observed
close to the symmetry axis. Fig.~\ref{fig:etet} shows the total
isotropic equivalent energy versus the off-axis angle. As underlined
by the linear-logarithmic scale, the jet profile is well described by
an exponential drop, especially if all the relativistic material is
considered. A very sharp jet/no-jet transition is instead observed if
only the hyper-relativistic material is singled out
($\Gamma_\infty\ge100$).

\section{Summary}

We have presented AMR special relativistic hydro simulations of a
relativistic light jet propagating through the core of a massive
star. We concentrate on the interaction of the jet with the stellar
material, showing that the emerging jet displays a rich phenomenology
of temporal and angular properties. We compute light curves and show
that they present some features observed in observations. In
particular, quasi isotropic cocoon outflow and the widening of the jet
opening angle can explain long dead times and precursors as observed
in BATSE and Swift light curves.

\acknowledgments
The software used in this work was in part developed by the
DOE-supported ASC/Alliance Center for Astrophysical Thermonuclear
Flashes at the University of Chicago.  This work was supported by NSF
grant AST-0307502, NASA Astrophysical Theory Grant NNG06GI06G, and
Swift Guest Investigator Program NNX06AB69G. This work was partially
supported by the National Center for Supercomputing Applications under
grant number AST050038 and utilized the NCSA Xeon cluster.

\end{document}